%% file: pos_wiese.tex
\documentclass{PoS}

\usepackage{graphicx,amsmath,amssymb,fontenc,times,mathptmx}

\usepackage[utf8]{inputenc}

\usepackage{nicefrac}

\usepackage{tikz}
\usetikzlibrary{shadows}
\usetikzlibrary{arrows,shapes,positioning}
\usetikzlibrary{decorations.markings}
\usetikzlibrary{decorations.pathmorphing}
\usetikzlibrary{decorations.pathreplacing}
\tikzstyle arrowstyle=[scale=1]
\tikzstyle directed=[postaction={decorate,decoration={markings, mark=at position .65 with {\arrow[arrowstyle]{stealth}}}}]
\tikzstyle end directed=[postaction={decorate,decoration={markings, mark=at position 1 with {\arrow[arrowstyle]{stealth}}}}]
\tikzstyle reverse directed=[postaction={decorate,decoration={markings, mark=at position .65 with {\arrowreversed[arrowstyle]{stealth};}}}]
\tikzstyle{ann} = [fill=white,font=\footnotesize,inner sep=1pt]

\title{Recent results for the proton spin decomposition from
lattice QCD}

\ShortTitle{Proton spin from the lattice}

\author{Constantia Alexandrou$^{ab}$,  
        Martha Constantinou$^{ab}$,  Kyriakos
	Hadjiyiannakou$^{c}$, Christos
	Kallidonis$^{b}$, Giannis Koutsou$^{b}$, 
	Karl Jansen$^{d}$, Haralambos Panagopoulos$^{a}$, Fernanda Steffens$^{d}$,
	Alejandro Vaquero$^{e}$,
	\speaker{Christian Wiese}$\;^{d}$\\
\\
        \llap{$^a$}Department of Physics, University of Cyprus, P.O. Box 20537, 1678 Nicosia, Cyprus\\
	\llap{$^b$}Computation-based Science and Technology Research Center, Cyprus Institute, 20 Kavafi Str., Nicosia 2121, Cyprus\\
	\llap{$^c$}Department of Physics, The George Washington University, Washington, DC 20052, USA\\
	\llap{$^d$}John von Neumann Institute for Computing (NIC), DESY, Platanenallee 6, D-15738 Zeuthen, Germany\\
	\llap{$^e$}INFN Sezione di Milano-Bicocca, Edificio U2, Piazza della Scienza 3, 20126 Milano, Italy

	\\
        E-mail: \email{christian.wiese@desy.de}}

\abstract{The exact decomposition of the proton spin has
been a much debated topic, on the experimental as well as
the theoretical side. In this talk we would like to report
on recent non-perturbative results and ongoing efforts to
explore the proton spin from lattice QCD. We present results
for the relevant generalized form factors from gauge field ensembles that
feature a physical value of the pion mass. These generalized form
factors can be used to determine the total spin and 
angular momentum carried by the quarks. In addition
we present first results for our ongoing effort to compute
the angular momentum of the gluons in the proton.}

\FullConference{XXIV International Workshop on Deep-Inelastic Scattering and Related Subjects\\
		11-15 April, 2016\\
		DESY Hamburg, Germany}

\begin{document}

\section{Introduction}
The understanding of the proton spin decomposition is an important topic within the nucleon structure
community. In particular, this is motivated by the
experimental measurement and analysis of the spin dependent
structure function of the proton $g_1^p$ by the European
Muon Collaboration \cite{Ashman:1989ig}, which showed that
the quark spin only contributes about half the value of the
proton's total spin.
Usually, these quantities are extracted from spin dependent structure functions which are
obtained from polarized scattering experiments as they have
for instance been performed at SLAC, JLab, CERN and DESY.
Unfortunately, there are not as many data points available as
for unpolarized DIS and consequently the
uncertainties for the extracted structure functions,
especially for the gluon, are rather large (see {\it e.g.}
Ref.\,\cite{Adolph:2015saz}).

Thus, an alternative determination of these quantities from first
principles would certainly be most useful and moreover provide a
rigorous test to the theory of QCD. For this task lattice
QCD is a promising candidate, since it is able to make {\it
ab initio} prediction for QCD observables in various energy
regimes. This includes matrix elements of local operators,
which are an essential tool for studying nucleon structure.
In particular these matrix elements can be related to the
form factors relevant for the nucleon spin and thus have the
potential to reveal the spin structure from a first principle
lattice calculation.

\section{The proton spin}
One gauge invariant possibility to decompose the proton spin
was proposed in Ref.\,\cite{Ji:1996ek} and is widely known
as Ji's sum rule. It decomposes the proton spin in the
following parts
\begin{equation}
  \frac{1}{2}=\frac{1}{2}\Delta \Sigma + L_q + J_g\,,
\end{equation}
with the intrinsic quark spin $\frac{1}{2}\Delta \Sigma$,
the total quark orbital angular momentum $L_q$ and the gluon
angular momentum $J_g$. The sum of the first two give the
total angular momentum of all quarks $J_q =
\frac{1}{2}\Delta \Sigma + L_q$. Unlike $J_q$, $J_g$ cannot
be further decomposed in a gauge invariant way.
The quark angular momentum can be expressed as the sum of two
proton form factors
\begin{equation}
  J_q = \sum_q \frac{1}{2}(A_{20}^q(0) + B_{20}^q(0))
\end{equation}
which are defined as moments of certain generalized parton
distributions (GPDs)
\begin{equation}
  A_{20}^q(0) = \int_{-1}^1 dx\,x \, H^q(x,0,0) = \langle x
  \rangle_q\,,\hspace{1cm} B_{20}^q(0) = \int_{-1}^1 dx\,x \,
  E^q(x,0,0)\,.
\end{equation}
where $\langle x \rangle_q$ is the average quark momentum
fraction in the proton.
A review on the various GPDs can be found in Ref.\,\cite{Ji:1998pc}.
The gluon angular momentum $J_g$ can be expressed
correspondingly.
The quark spin $\frac{1}{2}\Delta \Sigma$ can be related to
the proton singlet axial charge
\begin{equation}
  \Delta \Sigma = g_A^{(0)} = \sum_q \tilde{A}_{10}^q(0) =
  \sum_q \int_{-1}^1 \tilde{H}^q(x,0,0)\,.
\end{equation}
From the phenomenological analysis of experimental data
there are estimates of parts of the spin decomposition for instance  
for the quark spin $0.13 < \frac{1}{2} \Delta \Sigma < 0.18$
\cite{Adolph:2015saz} or the total quark angular momentum
$0.24 < J^q < 0.30$ \cite{Kroll:2014tma}. 

\section{Lattice setup}
On the lattice we can extract form factors from proton matrix
elements of certain operators $\langle P \vert \mathcal O
\vert P \rangle$ \cite{Abdel-Rehim:2015owa}. For the spin decomposition these are
the local axial-vector and one-derivative vector operators
\begin{equation}
  \mathcal O_A^{\mu,q} = \bar{q} \gamma^{\mu} \gamma^5 q
  \hspace{1cm} O_V^{\mu\nu,q} = \bar{q}
  \gamma^{\{\mu}D^{\nu\}} q\,
\end{equation}
with $D^{\nu}$ being the covariant forward and backward
derivative and $\{\hdots\}$ representing symmetrization and
subtraction of trace.

There are the following relation between these matrix
elements and the relevant form factors
\begin{equation}
  \langle P \vert \mathcal O_A^{\mu,q} \vert P
  \rangle =  \bar{u}(P) \tilde{A}_{10}^q(0)
  \gamma^{\mu}\gamma^5 u(P)
\end{equation}
\begin{equation}
  \langle P \vert \mathcal O_V^{\mu\nu,q} \vert P'
  \rangle =  \bar{u}(P) \left ( {A}_{20}^q(t)
  \gamma^{\{\mu}\overline{P}^{\nu\}} + {B}_{20}^q(t)
  \frac{\sigma^{\{\mu \rho} \Delta_{\rho}
  \overline{P}^{\nu\}}}{2m}+
{C}_{20}^q(t) \frac{\Delta^{\{\mu} \Delta^{\nu\}}}{m}\right)
u(P)\,
\end{equation}
where $\overline{P}=\frac{P+P'}{2}$, $\Delta = P - P'$ and
$t=\Delta^2$. In the limit $P=P'$ only the ${A}_{20}^q$ form
factor remains.

\begin{figure}
  \centering
  \input{tikz_wick3pt.tex} \hspace{1cm}
  \input{tikz_wick3pt_dis.tex} \hspace{1cm} \input{tikz_wick3pt_gluon.tex}
  \caption{\label{fig_wick}Possible Wick contractions
  of three-point correlation functions for the spin
  decomposition. {\bf Left:} connected {\bf middle:} disconnected quark loop or {\bf right:} gluon loop.}
\end{figure}
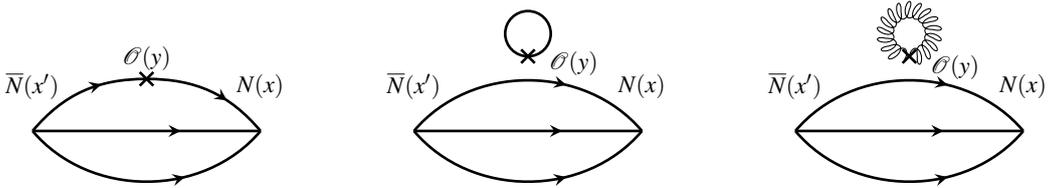

For the gluon content one can use the gluon energy
momentum tensor in order to extract the relevant form factor
\begin{equation}
  \langle P \vert \mathcal O_V^{\mu\nu,g} \vert P \rangle =
  \bar{u}(P) {A}_{20}^g(0)\gamma^{\{\mu} P^{\nu\}}
  u(P)\,,\hspace{.5cm}\text{with}\hspace{.5cm}O_V^{\mu\nu,g} = G^{\{\mu \rho}{G_{\rho}}^{\nu\}}\, 
  \label{eqn_gluon}
\end{equation}
where $G^{\mu\nu}$ is the field strength tensor. Certainly,
$B_{20}^g$ contributes to the gluon angular momentum as
well, but cannot be computed at the moment and is thus
neglected.
These matrix elements of operators can be related to the
ratio between three- and two-point correlation functions. The
correlation function is constructed from interpolating
fields of the proton and the operator
\cite{Abdel-Rehim:2015owa}. Free quark fields
have to be contracted to quark propagators using Wick's
theorem. Depending on the choice of the operator there are
different possibilities to perform the contractions. They
can be categorized into three general types of contractions,
as shown in Fig.\,\ref{fig_wick}.
All depicted types of contractions have to be considered in
order to explore the spin composition. While the here
considered light quark form factors will have connected and
disconnected contractions, one will find only quark
disconnected contractions for the strange and heavier quarks
and the gluon disconnected contraction for the gluon
contribution to the spin.

All results that will be presented in this work are computed
on the latest European Twisted Mass Collaboration (ETMC)
gauge field ensemble with a volume of $L^3 \times T = 48^3
\times 96$ \cite{Abdel-Rehim:2015pwa}. It features $N_f=2$
({\it i.e.} mass degenerate up and down quarks) dynamical fermions at
maximal twist, which ensures an improved continuum limit for
all here considered quantities \cite{Frezzotti:2003ni}. The
twisted mass parameter is set to $\mu=0.0009$ which
corresponds to a physical pion mass of $m_{\pi}\approx
133\,\text{MeV}$. A coupling of $\beta=2.1$ is used which
amounts to a lattice spacing of $a\approx 0.093\,\text{fm}$. 
The number of measurements and the source-sink separation
for the different form factors can be found in
Tab.\,\ref{tab_stat}. If the
chosen values for the source-sink separation are really sufficient
to suppress possible excited state effects has to be shown
in the future.

\section{Results for quark form factors}
The collected results for all available quark form factors
can be found in Table\,\ref{tab_qff}. This includes
connected and disconnected results for up, down and strange
quark form factors. A more elaborate discussion of the
computation of the connected part of the form factors from
the physical point ensemble can be
found in Ref.\,\cite{Abdel-Rehim:2015owa}, while the
disconnected parts are presented in Ref.\,\cite{Abdel-Rehim:2015lha}. The charm and heavier quarks are
neglected in this analysis. For the $B_{20}^q$ form factor
there are up to this moment no disconnected results
available. All necessary renormalization factors were
obtained from a non-perturbative lattice calculation, {\it
cf.} \cite{Abdel-Rehim:2015owa} and references therein.
\begin{table}
\centering
\begin{tabular}{ c c c c c }
\hline
 {$\Delta \Sigma$} & up & down & strange & combined\\
\hline
connected & 0.904(23) & -0.311(12) & - & 0.594(24)\\
\hline
disconnected & -0.076(16)& -0.076(16) & -0.042(10) &
-0.194(25) \\
\hline
combined & 0.828(32) & -0.387(20) & -0.042(10) &
{0.400(35)} \\
\hline
\hline
{$A_{20}(0)=\langle x \rangle$} & up & down & strange & combined\\
\hline
connected & 0.346(9) & 0.152(6) & - & 0.497(12)\\
\hline
disconnected & 0.112(70)& 0.112(70) & 0.092(41) & 0.315(106) \\
\hline
combined & 0.458(70) & 0.264(70) & 0.092(41) &
{0.812(107)} \\
\hline
\hline
{$B_{20}(0)$} & up & down & strange & combined\\
\hline
connected & 0.133(40) & -0.149(40) & - &
{-0.016(56)} \\
\hline
\end{tabular}
\caption{\label{tab_qff}Renormalized results for various
quark form factors including up, down and strange quarks for
connected and disconnected contractions.}
\end{table}
\begin{table}
\centering
   \begin{tabular}{ c c c c }
     \hline
   quantity & \# conf.& \# source pos.& source-sink sep.\\
     \hline
   {$\Delta \Sigma$} conn. & 542& 88& $14a$\\
     \hline
    {$\Delta \Sigma$} disc.  & 2137 & 100 & $14a$\\
     \hline
    {$A_{20}$} conn.& 580 & 16 & $16a$\\
     \hline
    {$A_{20}$} disc. light& 1219 & 100 & $14a$\\
     \hline
    {$A_{20}$} disc. strange& 2153 & 100 & $14a$\\
     \hline
    {$B_{20}$}& 425 & 16 & $12a$\\
     \hline
   \end{tabular}
   \caption{\label{tab_stat} Statistics and source-sink
   seperation for the computed quark form factors.}
 \end{table}
From these values we can extract the following values for
the proton spin decomposition. For the quark spin we find
\begin{equation}
  \frac{1}{2}\Delta \Sigma = 0.200(17) \hspace{.5cm} \text{
  with }\hspace{.5cm} \frac{1}{2}\Delta u = 0.414(16)\,,
  \hspace{.5cm}\frac{1}{2}\Delta d = -0.194(10)\,,
  \hspace{.5cm}\frac{1}{2}\Delta s = -0.021(5)\,.
\end{equation}
$\frac{1}{2}\Delta \Sigma$ is slightly above the boundaries that are given
from the phenomenological analysis of experimental data
$0.13 < \frac{1}{2} \Delta \Sigma < 0.18$ \cite{Adolph:2015saz}. 
For the total angular momentum
of all quarks we get the following value, where we have
neglected a possible disconnected contribution for the
$B_{20}^q$ form factor
\begin{equation}
  J_q = 0.398(60) \hspace{.5cm} \text{ with }\hspace{.5cm}
  J_u = 0.296(40)\,,\hspace{.5cm} J_d = 0.058(40)\,,
  \hspace{.5cm} J_s = 0.046(20)\,.
\end{equation}
As a notable fact the down quark barely contributes to the
total angular momentum, which is in a rather good agreement with the
analysis of GPDs \cite{Kroll:2014tma}, as are the results
for the other quark flavors. From these results one
can infer the values for the orbital angular momentum from
$L_q = J_q - \frac{1}{2}\Delta \Sigma$
\begin{equation}
  L_q = 0.198(62) \hspace{.5cm} \text{ with }\hspace{.5cm}
  L_u = -0.118(43)\,,
  \hspace{.5cm} L_d = 0.252(41)\,,
  \hspace{.5cm} L_s = 0.067(21)\,.
\end{equation}
This shows that the orbital angular momentum of all quarks
is of the order of the quark spin.

\section{Results for the gluon content}
On the basis of Eq.\,(\ref{eqn_gluon}) we can try to compute the
$A_{20}^g$ form factor on the lattice.
On the lattice the gluon operator can be expressed in terms
of plaquette terms, using $U_{\mu\nu} =
\exp(iga^2G_{\mu\nu}+O(a^3))$. We choose a representation
where the gluon form factor can be extracted without
applying a momentum boost to the proton.
\begin{equation}
  \mathcal O_V^g=
  \frac{2}{9}\frac{\beta}{a^4}\left(\sum_i\mathrm{Re}(U_{i4})-\sum_{i<j}\mathrm{Re}(U_{ij})\right)\,.
\end{equation}
From the correlation of this operator with the nucleon
two-point function, as depicted on the right side of
Fig.\,\ref{fig_wick}, we were able to extract the gluon form
factor. Up to 20 steps of 4D stout smearing \cite{Morningstar:2003gk}
with $\omega = 0.1315$ were used in order to remove
the fluctuations due to the gauge field and thus obtain a
statistically significant signal. We extract a signal from
several values for the source-sink separation and obtained a
bare lattice result from a combined plateau fit to these
values
\begin{equation}
  {A_{20}^g}_{\text{,bare}} = \langle x
  \rangle^g_{\text{bare}} =
  0.318(24)\hspace{.5cm}\xrightarrow{\mathrm{renormalization}}
  \hspace{.5cm}A_{20}^g =
  0.321(25)\,.
\end{equation}
Since the gluon operator is a singlet operator it mixes with
other singlet operators, most importantly the quark singlet
operator ${A_{20}^q}_{\text{,bare}}$. The thus necessary
renormalization and mixing coefficients were computed in a
one-loop perturbative lattice calculation that also takes
stout smearing into account and will be presented in an
upcoming paper. 
For the $B_{20}^g$ form factor we are not yet able to present any
results.

\section{Conclusion and Outlook}
\begin{figure}
  \centering
  \includegraphics[scale=0.35]{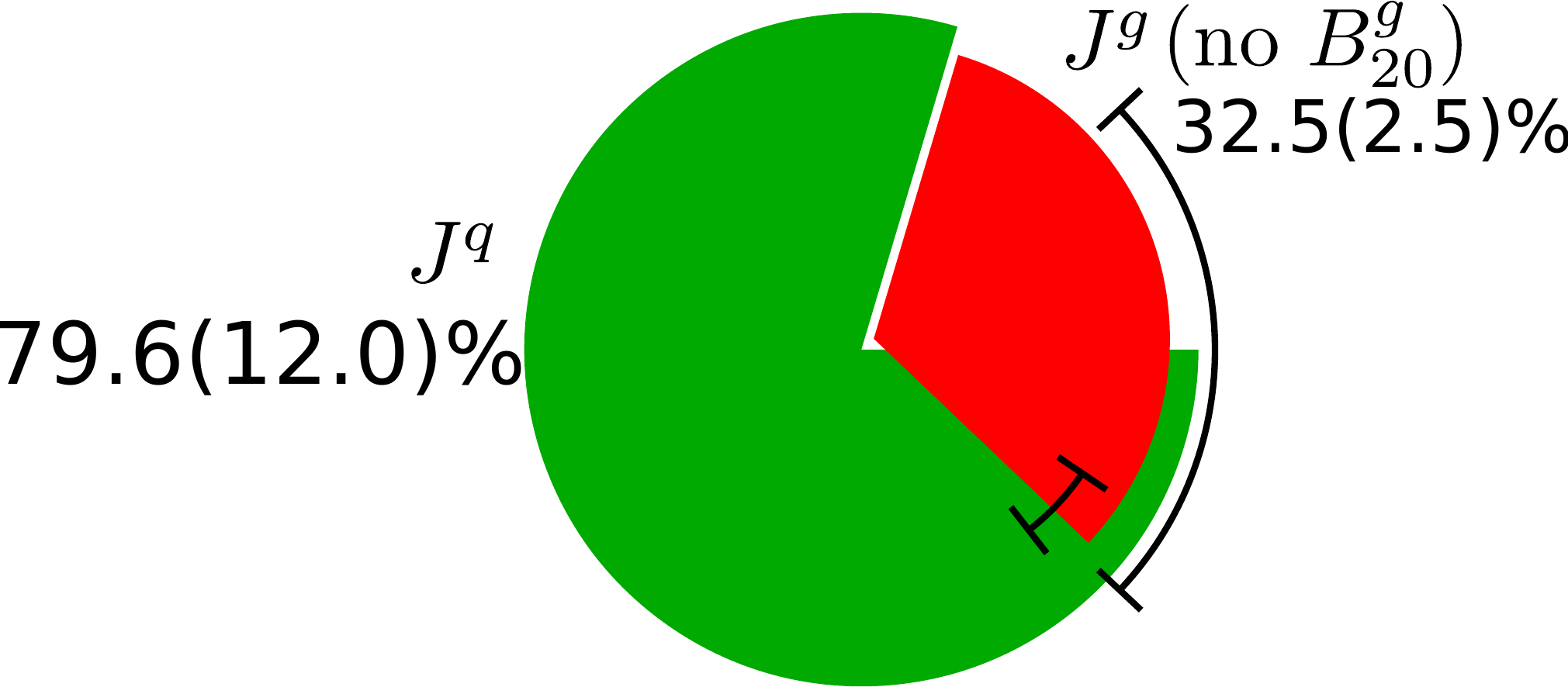}\hspace{.1cm}\includegraphics[scale=0.35]{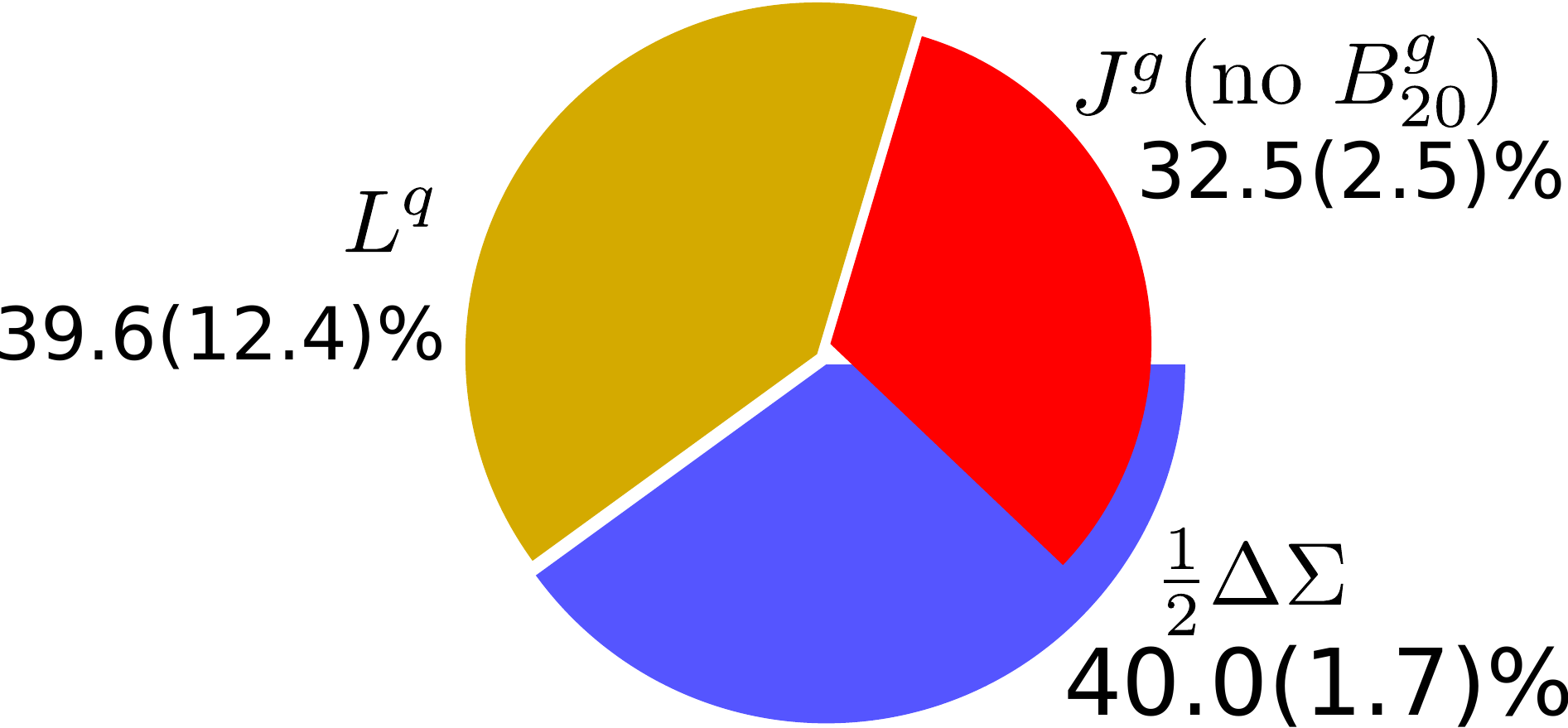}
  \caption{\label{fig_sum}Graphical representation of the
  proton spin contribution of quarks and gluons from lattice QCD
results.}
\end{figure}

Taking all current results into account, we present a first preliminary lattice analysis for the proton spin
decomposition. This certainly is not a final lattice result
since systematic effects, {\it e.g.} cut-off effects, have
not thoroughly been studied and it was not possible to
obtain results for all the necessary form factors. A first
test of our results is the reconstruction of the
total proton spin from lattice data from the individual form
factors as shown in Fig.\,\ref{fig_sum}.
At the moment we obtain about 110 percent of the total
nucleon spin, however, with a large uncertainty of $J_q$,
as indicated by the errorbars in the chart. 
If this is really a statistical effect or caused by
systematic uncertainties or missing parts of the form
factors is something to be explored in further studies.
Nevertheless, the here presented results already provide a
good qualitative study of the proton spin decomposition and
certainly show that it is feasible to explore this important
topic from a lattice point of view. Especially the fact
that all results are extracted from a gauge ensemble with a
physical pion mass will allow us to focus on further
reducing other systematic effects and get closer to a good
prediction for the proton spin. 
If this approach proves to be successful it could be easily
generalized to other bound states and might in the future
enable us to explore the spin structure of many hadrons.

\bibliographystyle{JHEP}
\bibliography{references.bib}

\end{document}

%% file: tikz_wick3pt.tex
\begin{tikzpicture}[scale = 0.3]

\node[ann] at (10,2) {$N(x)$};
\node[ann] at (0,2) {$\overline{N}(x')$};
\node[ann] at (5,3.2) {$\mathcal O(y)$};

\draw[black, directed, line width=1pt] (0,0) -- (10,0);
\draw[black, directed,line width=1pt]  (5.0,2.3) to[out=0,in=130] (10,0);
\draw[black,directed, line width=1pt] (0,0) to[out=50,in=180]   (5,2.3);
\draw[black, directed,line width=1pt] (0,0) to[out=-50,in=-130] (10,0);
\draw[black, line width=1pt] (4.7,2.6) -- (5.3,2.0);
\draw[black, line width=1pt] (5.3,2.6) -- (4.7,2.0);
\end{tikzpicture}

%% file: tikz_wick3pt_dis.tex
\begin{tikzpicture}[scale = 0.3]

\node[ann] at (10,2) {$N(x)$};
\node[ann] at (0,2) {$\overline{N}(x')$};
\node[ann] at (7,3) {$\mathcal O(y)$};

\draw[black, directed, line width=1pt] (0,0) -- (10,0);
\draw[black,directed, line width=1pt] (0,0) to[out=50,in=130]   (10,0);
\draw[black, directed,line width=1pt] (0,0) to[out=-50,in=-130] (10,0);

\draw[black, line width=1pt] (4.7,3.6) -- (5.3,3.0);
\draw[black, line width=1pt] (5.3,3.6) -- (4.7,3.0);

\draw[black, line width=1pt] (5.0,4.3) circle(1cm);

\end{tikzpicture}

%% file: tikz_wick3pt_gluon.tex
\begin{tikzpicture}[scale = 0.3]

\node[ann] at (0,2) {$\overline{N}(x')$};
\node[ann] at (10,2) {$N(x)$};
\node[ann] at (7,2.8) {$\mathcal O(y)$};

\draw[black, directed, line width=1pt] (0,0) -- (10,0);
\draw[black,directed, line width=1pt] (0,0) to[out=50,in=130]   (10,0);
\draw[black, directed,line width=1pt] (0,0) to[out=-50,in=-130] (10,0);

\draw[black, line width=1pt] (4.7,3.6) -- (5.3,3.0);
\draw[black, line width=1pt] (5.3,3.6) -- (4.7,3.0);

\draw[decorate, rotate
around={-90:(5.0,4.3)},decoration={coil,amplitude=3pt,
segment length=3pt}] (5.0,4.3) circle(1cm);

\end{tikzpicture}